# Boundary-Induced Embedded Eigenstate in a Single Resonator for Advanced Sensing

**AUTHORS:** Rasmus E. Jacobsen[1], Alex Krasnok[2], Samel Arslanagić[3], Andrei V. Lavrinenko[1*], and Andrea Alù[2,4*]

**AFFILIATIONS:**

[1]Department of Photonics Engineering, Technical University of Denmark, Bld. 345A, Ørsteds Plads, 2800 Kgs. Lyngby, Denmark (e-mail: rajac@fotonik.dtu.dk, alav@fotonik.dtu.dk)

[2]Photonics Initiative, Advanced Science Research Center, City University of New York, New York, NY 10031, USA (akrasnok@gc.cuny.edu, aalu@gc.cuny.edu)

[3]Department of Electrical Engineering, Technical University of Denmark, Bld. 348, Ørsteds Plads, 2800 Kgs. Lyngby, Denmark (e-mail: sar@elektro.dtu.dk)

[4]Physics Program, Graduate Center, City University of New York, New York, NY 10016, USA

*Corresponding author: e-mail: alav@fotonik.dtu.dk, aalu@gc.cuny.edu

**ABSTRACT**

Electromagnetic embedded eigenstates, also known as bound states in the continuum (BICs), hold a great potential for applications in sensing, lasing, enhanced nonlinearities and energy harvesting. However, their demonstrations so far have been limited to large-area periodic arrays of suitably tailored elements, with fundamental restrictions on the overall footprint and performance in presence of inevitable disorder. In this work, we demonstrate a BIC localized in a single subwavelength resonator obtained by suitably tailoring the boundaries around it, enabling a new degree of control for on-demand symmetry breaking. We experimentally demonstrate how boundary-induced BICs open exciting opportunities for sensing by tracing the dissolution of NaCl in water, determining evaporation rates of distilled and saltwater with a resolution of less than 1 µL using a tabletop experimental setup.

**ONE SENTENCE SUMMARY**
We realized a localized embedded eigenstate in a single resonator by suitably controlling the boundaries around it, and explored its operation for advanced sensing and tracing of chemical reactions.

**MAIN TEXT**
Embedded eigenstates, also known as bound states in the continuum (BICs), have recently become an area of great interest in wave physics. They were predicted in the seminal work by von Neumann and Wigner as a curious eigen-solution of single-particle Schrödinger equation that can reside within the continuum despite being compatible with decay in terms of momentum (*1*). Given the mathematical analogies with the wave equation in electrodynamics, optics and linearized acoustics, this phenomenon has been experimentally observed in several wave settings (*2*). Theoretically speaking, a BIC is a resonant state of an open system that supports an infinite quality factor (Q-factor), despite supporting a momentum compatible with radiation (*3*). Due to this fascinating property, BICs offer exciting prospectives for various applications, including sensing (*4*), lasing (*5–7*), nonreciprocity (*8*), enhanced thermal emission (*9*, *10*) and energy harvesting (*11*).

BICs can be classified according to the number of eigenmodes involved in the scattering process (single or multiple) and to the nature of the underlying interference mechanisms (symmetry-enabled or accidental). The simplest example of nonradiative, symmetry-enabled, single-mode BIC in an open system is a hedgehog-like collection of dipoles longitudinally arranged over a sphere or a charged pulsating spherical shell, for which radiation is forbidden due to symmetry (*12*). This non-radiating current distribution was employed as a first attempt to explain the stability of atoms (*13-16*). While this idealized model is practically challenging to implement, BICs have been demonstrated in periodic 2D arrays of dipoles that do not radiate due to mutual destructive interference. Accidental single-mode BICs can also be supported by multilayers involving epsilon-near-zero (ENZ) or other extreme material parameters (*10*, *17–19*). Friedrich and Wintgen predicted BICs in systems involving multiple, not orthogonal coupled modes (*20*). These BICs leverage destructive interference of at least two resonant modes coupled to the same radiation channel, and they can exist in both symmetry-protected and accidental form. This approach has been of great interest in the recent literature, as it enables nonradiative states tailored by variations of the system parameters, e.g., angle of incidence or refractive index, becoming particularly attractive in various photonic crystal setups (*2*, *21–27*).

In this context, the formation of a symmetry-protected BIC at the Γ-point of a periodic array is schematically illustrated in Fig. 1. We first consider an individual resonator, in the form of a metallic cylindrical disk placed over on a conducting ground plane, which resonantly scatters light as a vertical electric dipole (*28*) around the frequency of interest, Fig 1A. When illuminated by an incident wave, the scattering pattern has a null along the dipole, and the resonant spectrum is characterized by a Q-factor controlled by the power scattered towards all other directions into the radiation continuum. Next, we consider a periodic array of such resonators: choosing a subwavelength periodicity, the scattered power is directed towards the zeroth diffraction order (except at the ends of the array), i.e., a single direction, due to destructive interference among the array elements for all other angles. By aligning this outgoing diffraction channel with the null in the scattering pattern of each resonator, a symmetry-protected BIC emerges at the Γ-point of the first Brillouin zone, i.e., in the direction normal to the array (*2*, *3*, *5*, *29*). In this scenario, the resonant mode sustains vertical dipoles all with the same amplitude and phase, as sketched in Fig. 1B, which cannot radiate towards the normal. Because of reciprocity, the periodic array cannot be excited by a plane wave at normal incidence, hence forming a nonradiating mode unobservable in the scattering spectra. Coupling to outgoing radiation can be triggered by small broken symmetries along the array, resulting in a sharp Fano line-shape in the radiation spectrum often referred to as a quasi-BIC (*27*, *30*, *31*). These phenomena can be observed in large-area arrays, implying large footprints and challenges in ensuring these precise symmetries across large areas. More importantly, these features are not well suited for sensing: the BIC mode is extended in space, and therefore the asymmetries controlling the coupling to radiation need to be equally extended. For similar reasons, despite the very large Q factors, these BICs do not support large local density of states (LDOS) *(17)*.

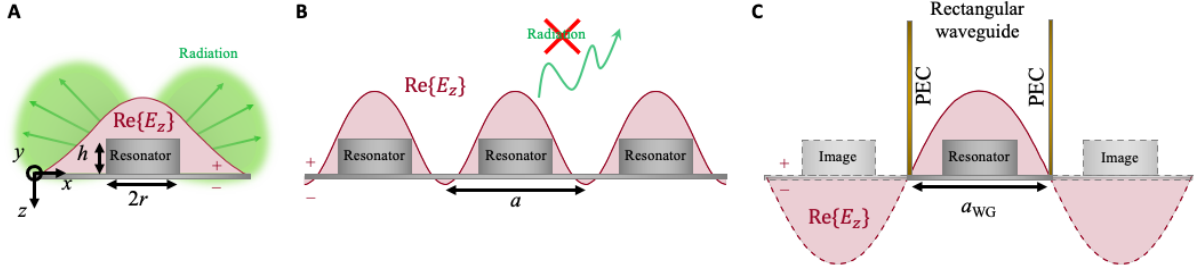

**Fig. 1. Concept of an extended BIC in a periodic array and a boundary-induced localized BIC in a single resonator.** (**A**) Sketch of a single metallic resonant disk over a ground plane, along with the *z*-component of the E-field distribution of its vertical dipole mode. The parameters are $h = 25$ mm, $r = 30$ mm, $f_r \approx 1{,}940$ MHz. The corresponding Q-factor of the eigenmode is ~2. (**B**) Periodic array of metallic resonators. An infinite array supports an extended Γ-point BIC when all vertical dipole moments are excited with the same amplitude and phase. This Γ-point BIC turns into a quasi-BIC in the case of a finite structure and/or the presence of asymmetries or material loss. The corresponding near-field distribution is shown in red, and the calculated Q-factor of the eigenmode is ~21,000. The parameters are $h = 25$ mm, $r = 30$ mm, $a = 110$ mm, $f_r \approx 2{,}470$ MHz. (**C**) Single metallic resonator in a rectangular waveguide: even though we use a single resonant element, the structure can support a localized BIC. If the waveguide walls and the metallic resonator are not perfectly conductive, the BIC turns into a quasi-BIC. In our case, the calculated Q-factor of the eigenmode is ~8,000, which can be increased using lower loss materials. The parameters are $h = 27$ mm, $r = 18$ mm, $a_{WG} = b_{WG}/2 = 54.61$ mm, $f_r \approx 1{,}890$ MHz. All $E_z(x)$ are taken at 20 mm above the disks' top base. The lines above/below the ground plane correspond to positive/negative real values of $E_z(x)$.

The BICs demonstrated to date have been based on these principles, hence they emerge in spatially extended and bulky periodic structures, hindering their implementation and effective use as sensors or other forms of enhanced light-matter interactions. In contrast to these structures, localized BICs in single resonators have been predicted to offer extremely large LDOS, ideal for sensing applications *(32, 33)*. However, to date localized BICs have been leveraging the use of extreme material parameters, making them prone to losses and challenging in their practical implementation. Friedrich-Wintgen supercavity modes in high-index dielectric nanoparticles are intrinsically open to radiation, and therefore they cannot support true BICs even in theory *(34)*.

In this work, we explore a different approach to enable localized BICs. We suitably modify the boundary conditions around a single resonator in order to control the destructive interference between the resonator and its images induced by the tailored boundaries, thus inducing nonradiating resonant modes. In particular, we can place the individual resonant disk at a suitable location within two lateral metallic walls, Fig. 1C. The resonator, excited by the fundamental TE$_{10}$ mode of the resulting waveguide, interacts with its images created by the lateral walls, assuming that they are good conductors. The result is an effectively 2D periodic array with alternating field signs, as sketched in Fig. 1C and further analyzed in Fig. S1-S3. This opens the opportunity to induce a localized BIC, controlled by the resonator geometry and the distance between the lateral walls, confining the extended BIC in the periodic array of Fig. 1B into a single resonant element. A tailored asymmetry introduced in our resonator is automatically extended to the image array, offering a powerful tool to tailor the Q-factor at will, as envisioned in *(27)*, but without having to control an extended broken symmetry across the entire array.

We control the asymmetry in our structure by introducing a tiny water droplet placed atop the metallic resonator, leveraging the high permittivity of water at microwaves to trigger non-negligible perturbations in the local fields (*36–38*). Different from the case of extended BICs, here the LDOS in the resonator is largely enhanced at the BIC condition, with a controllable Q-factor that enables sensing of deeply subwavelength variations to the water volume, as well as small changes in its refractive index. Remarkably, as we show in the following, we can use this platform to trace in real time the occurrence of chemical reactions, e.g., the dissolution of NaCl in water. Similarly, we can determine the evaporation rates of distilled and salt water with a resolution of less than 1 µL.

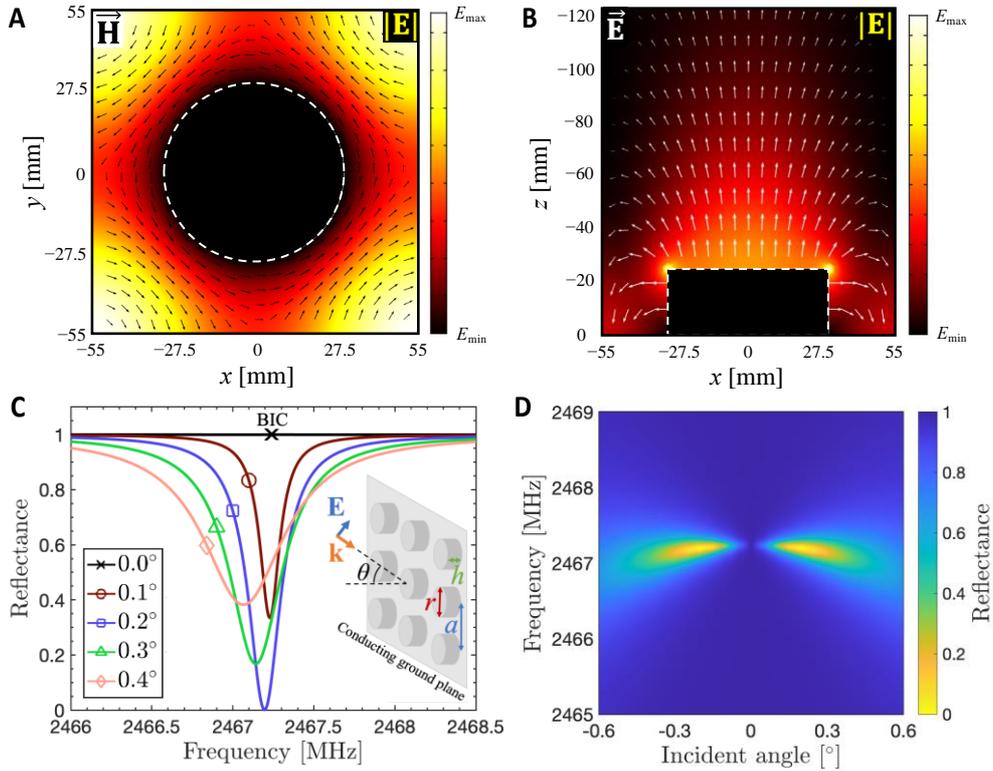

**Fig. 2. Symmetry-protected extended BIC in a periodic array.** (A) and (B) Magnitude of the total E-field [V/m] (colorbars) of the eigenmode in the *xy*- and *yz*-plane. (C) and (D) Numerically calculated reflectance spectra for different incident angles. The black and white arrows show the instantaneous magnetic and electric fields, respectively, both in a logarithmic scale. The geometric parameters are $r = 30$ mm, $h = 25$ mm and $a = 110$ mm, and the intrinsic metal conductivity is $\sigma = 5.8 \times 10^7$ S/m corresponding to that of copper. The calculated real eigenfrequency is 2,467 MHz, and the Q-factor using realistic metal loss is 21,000.

**Symmetry-protected extended BIC**

In order to outline the physical mechanisms supporting our boundary-induced localized BIC, in Fig. 2 we first numerically analyze the conventional extended BIC supported by a periodic array of resonant metallic disks backed by a conducting ground plane, as in Fig. 1B (for details of our analysis, see *Supplementary Materials*). For the chosen geometrical parameters, an eigenmode analysis of the array predicts an extended BIC arising at the frequency of 2,467 MHz, with an unbounded Q-factor in the ideal scenario of perfect electric conductors (PEC). Considering a realistic metal conductivity, the calculated resonance Q-factor is 21,000. Figs. 2 A and B show the magnitude of the E-field distribution of the corresponding eigenmode in the cross-sections of one of the resonators. The disk supports a high-intensity electric field on its top side, decaying away from the disk. The circulating magnetic field and vertical electric field distributions, plotted as white and black arrows in Fig. 2, A and B, respectively, confirm that the resonance is dominated by a vertical electric dipole component. Standing wave patterns arise in the *xy*-plane due to the coupling with neighboring resonators.

By illuminating the array with a *p*-polarized (i.e., parallel to the plane of incidence) plane wave at different incident angles $\theta$, the reflection spectra (Fig. 2, C and D) showcase the emergence of a BIC. At any incident angle other than 0°, the incident wave strongly couples to the array resonance, inducing an absorption dip. For normal incidence, the wave is fully reflected and cannot couple to the resonant dip. With the considered level of metal loss, critical coupling and full absorption is achieved at 0.2°, for which the radiated power from the dipoles equals the absorbed power in the metal. At larger (smaller) incident angles, the array is over (under) coupled. Approaching normal incidence, the resonant linewidth gets narrower and narrower, until it vanishes because of symmetry at 0°, when the resonant mode becomes a BIC and cannot be excited by the incoming wave.

**Symmetry breaking**

By adding a tailored asymmetry in the array, we can control the coupling of this nonradiating mode to normally incident waves. We achieve symmetry breaking by adding a drop of distilled water [permittivity model taken from (*39*)] on the top side of each metallic resonator. The drops are modeled as hemi-spherical particles with radius $r_w$, and their center is displaced from the disk center by $(x, y) = (d_w, 0)$. By varying the drop size and position in each array element, we can control the coupling to a normally incident *x*-polarized plane wave, as shown in Fig. 3. In Fig. 3A, we fix the drop position, so that there is a gap of 1 mm between the drop and the resonator boundaries ($d_w = r − r_w − 1$ mm). We observe no change in reflection for $r_w = 2$ mm, due to the finite spectral resolution of our simulations. As more water is added, a narrow reduction in reflection can be observed, and critical coupling is reached with $r_w = 4.8$ mm (96.5 µL). A further increase in drop size results in a significant broadening of the reflection spectrum due to over-coupling. In the under-coupled regime, the resonance frequency is red-shifted as more water is added. Furthermore, the losses in water reduce the Q-factor from 21,000 (no water) to 4,300 for a drop size $r_w = 4.8$ mm.

In Fig. 3B we study variations in the drop position, where the drop size is fixed at $r_w = 4.8$ mm. When $d_w = 0$, the configuration is symmetric, and therefore the BIC is preserved. As the drop moves away from the center of the metallic disk, the incident wave couples to the array, and at $d_w = 24.2$ mm, the coupling is critical. As the drop approaches the edge of the resonator, the proximity effects cause the resonance frequency to change from blue-shifted to red-shifted. In Fig. 3C and D, the magnitude of the total E-field in the *xy*- and *xz*-planes are shown for a drop with $r_w$ = 4.8 mm and $d_w$ = 24.2 mm. The similarities between the field distributions in Fig. 2 and Fig. 3 confirm that the BIC mode is only slightly perturbed by adding water, and it can be excited despite the excitation at normal incidence due to the broken symmetry.

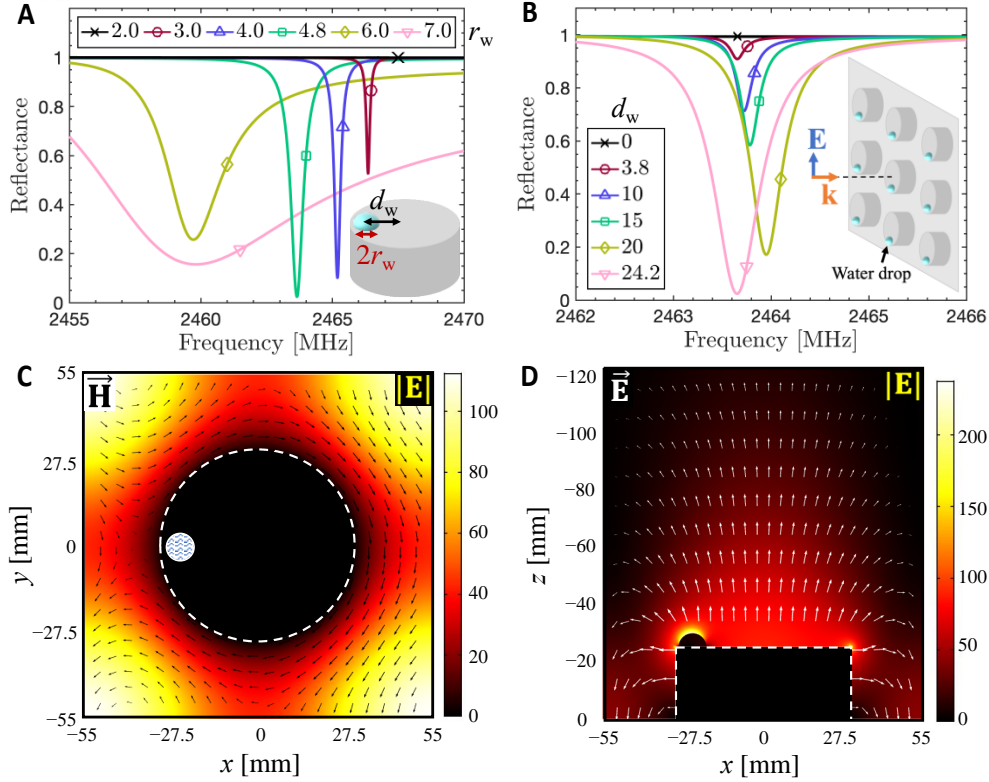

**Fig. 3. Symmetry breaking in an extended BIC**. (**A**) Numerically calculated reflectance spectra for different water drop sizes with $d_w = r - r_w - 1$ mm (legend shows the hemisphere radius in mm). (**B**) Reflectance spectrum for different water drop positions. The drop radius is $r_w$ = 4.8 mm (legend shows the displacement $d_w$ in mm). (**C**) and (**D**) Magnitude of the total E-field [V/m] (colorbars) in the *xy*- and *xz*-plane for a plane wave at normal incidence with $r_w$ = 4.8 mm and $d_w$ = 24.2 mm. Black and white arrows show the instantaneous magnetic and electric fields, respectively, both in logarithmic scale. The Q-factor is 4,300 for water drops with $r_w$ = 4.8 mm and $d_w$ = 24.2 mm. The water temperature is 20 °C in all simulations.

**Boundary-induced localized BIC**

The results in Fig. 3 show how it is possible to couple an extended BIC mode to a normally incident plane wave by introducing an array of symmetry-breaking defects in each element. Arguably this is practically challenging, and not amenable to a sensing platform, in which the sensor is typically localized. We overcome this challenge by considering a boundary-induced localized BIC, as introduced in Fig. 1C, confined into a single resonator placed in a rectangular metallic WR-430 waveguide. Our design dimensions are adapted to the WR-430 waveguide, with $r$ = 18 mm and $h$ = 27 mm, corresponding to a BIC eigenfrequency around 1,880 MHz according to a numerical eigenmode analysis (Fig. S4). A photograph of the resonator inserted in the waveguide section is shown in Fig. 4A. Both the resonator and ground plane are made of aluminum, and a small water container consisting of a cylindrical shell of Rohacell 51 HF is glued on the resonator. The low permittivity of Rohacell, measured in-house to be 1.075 (*36*), has a negligible effect on the response. Fig. 4B shows photographs of the resonator with different water amounts. The reflection coefficient is measured with a vector network analyzer (VNA), and the full experimental setup is sketched in Fig. 4C.

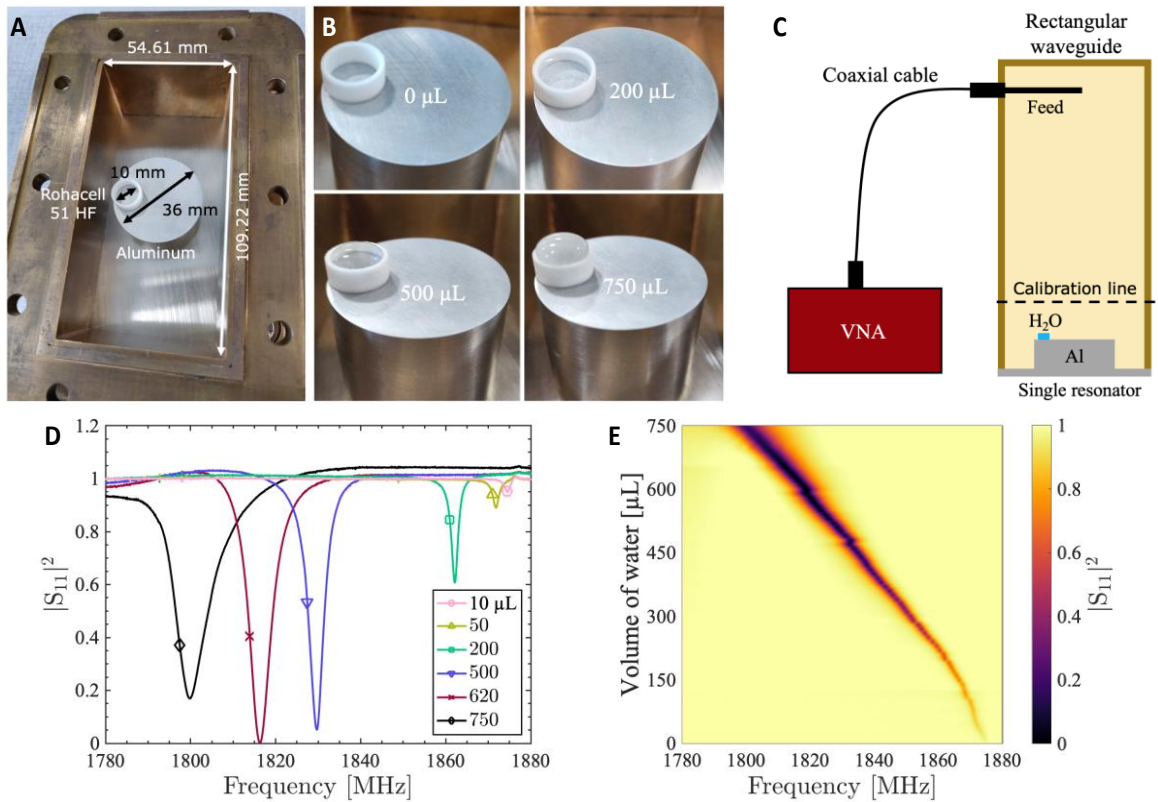

**Fig. 4. Localized BIC in a metallic waveguide.** (**A**) Photograph of a single disk resonator inserted in a WR-430 waveguide section. (**B**) Photographs of the resonator with different water fillings. (**C**) Sketch of the experimental setup. (**D**) Normalized reflection coefficient as a function of frequency and water volume. The water temperature is 24 °C. (**E**) Evolution of the quasi-BIC resonance evolving towards the BIC at ~1,880 MHz as water evaporates from the system (see Fig. S5C for the reflection phase).

The addition of a single water droplet in our localized boundary-induced BIC enables controlled symmetry breaking, hence leading to an efficient way to control the coupling of the BIC mode to the continuum of propagating modes in the waveguide. The results summarized in Fig. 4D show the frequency dependence of the measured reflection coefficient for different water volumes. The results have been normalized to the spectra of the resonator without water (for details, see *Supplementary Materials,* Fig. S5A). No resonance is excited with 0 µL water, since the system supports a nonradiating localized BIC, whereas with 10 µL the mode evolves to a quasi-BIC arising in the measured spectrum around 1,875 MHz. Adding more water increases the intensity of the resonance and shifts it towards lower frequencies. From 10 to 610 µL, we are in the under-coupled regime, and at 620 µL we hit critical coupling, for which the reflection coefficient is reduced to near-zero. Above 620 µL, the resonator is over-coupled and we observe an increase in reflection and resonance broadening. The frequency shift is directly proportional to the water volume from around 200 µL. Below 200 µL, the water is not evenly distributed in the Rohacell container due to surface tension, and therefore its volume shape changes as more water is added.

The system is ideally suited to observe in real time water evaporation from the system: Fig. 4E demonstrates how the quasi-BIC evolves towards the localized BIC as water is removed. The Q-factor as a function of water volume (see Fig. S5E) displays a steady growth as the water volume is reduced, due to the system approaching the BIC and the reduced water losses. At 10 µL, the Q-factor is ~1,500, but at the critical coupling regime (620 µL), the Q-factor decreases to 250.

**Sensing chemical reactions with a localized BIC**

Our results reveal that the resonance frequency linearly decreases with the water volume as ~0.1124 MHz/µL starting from 200 µL, Fig. 4E (see also Fig. S5B – D), allowing a precise determination of volume changes as low as 1 µL. We can use this tool to estimate the evaporation rates of distilled water and a NaCl solution, as shown in Fig. 5. Since the Rohacell container is open, water evaporates over time, causing a shift in resonance frequency ($\Delta f_{min}$) and of the reflection minimum, $(|S_{11}|^2)_{min}$. Figs. 5, A and B, demonstrate the dependence of $|S_{11}|^2_{min}$ and $\Delta f_{min}$ in function of time. From the curve slopes, we can carefully estimate the evaporation rate, and the estimated evaporated water volume as a function of time, shown in Fig. 5C.

We can also exploit the localized BIC to monitor in real time the effect of salination in water. The addition of a salt crystal triggers a rapid change in the response (see Fig. 5, A and B, period I). The NaCl density is lower when dissolved in water, thus the volume expands and we observe a red-shift in the resonance frequency. Hereafter (period II), the resonance frequency is blue-shifted as $Na^+$ and $Cl^-$ ions are dispersed in water (hydration), effectively increasing the solution density. Furthermore, the conductivity increases, thus increasing the reflection. After 20 minutes, the NaCl is fully dissolved (period III), and the slope is stabilized, such that we can estimate the evaporation rate.

The estimated evaporation rate is 213 nL/min for distilled water, whereas, for a 7.1 ppt NaCl solution, it is 201 nL/min, Fig. 5C, confirming that salt water evaporates more slowly than distilled water, since water molecules are attracted to the dissolved salt ions, and more energy is thus required to break them apart. We experimented with different sizes of NaCl crystals, Fig. 5D (see Fig. S6 for the frequency shift). Larger NaCl crystals change the response faster due to their size, but it also takes longer time to dissolve them, as seen by the slope changes. Even a small NaCl crystal of 0.3 mg (≈ 0.6 ppt in 500 µL water) can be clearly measured. Such a small change in salinity corresponds to an increase in the loss tangent of approximately 0.0066 [7.7 %, (40)].

There are several processes during the reaction that cause changes in volume and permittivity. A temperature shift changes both, as studied in Fig. S7 by simply heating the water before insertion. A temperature increase causes a decrease in density (i.e., larger volume) and permittivity, and therefore a red-shift and a blue-shift in frequency, respectively. In experiments, the frequency is blue-shifted, revealing that the volume change is the governing factor. This is interesting to observe, since only permittivity changes are typically considered in water-based structures (*37*),(*41*). In Fig. 5, E and F, we have isolated the permittivity change due to a variation in salinity by measuring NaCl solutions of equal volume of 500 µL. The reflectance (resonance frequency) increases (decreases) with increasing salinity due to the higher conductivity in the solution. The higher conductivity reduces the field penetration in the solution, and thus less power is absorbed. We measure changes with concentrations as low as 0.2 ppt (≈ 0.1 mg in 500 µL water), and from 0 to 5 ppt the change in response is linear with good approximation.

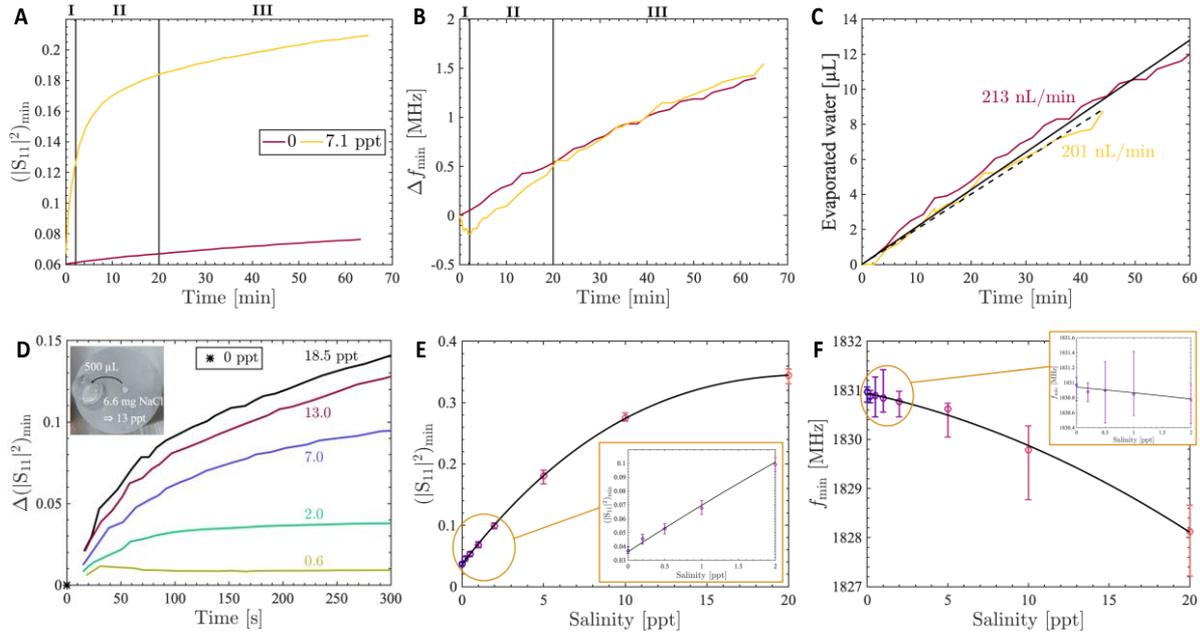

**Figure 5. Sensing and chemical reaction tracing with localized BIC.** (**A**), (**B**) and (**C**) $|S_{11}|^2_{min}$, $\Delta f_{min}$, and estimated evaporated water volume, respectively, as a function of time. Time 0 designates the point of the first measurement in (**A**) and (**B**), whereas it designates the point of full dissolvement of NaCl in water in (**C**). The black solid and dashed lines in (**C**) are the tendency lines of 0 ppt and 7.1 ppt solutions, respectively. (**D**) Chemical reaction of single NaCl crystals added to 500 µL of water. (**E**) and (**F**) show the response of 500 µL NaCl solutions with different salinities. (**D**) Change in the reflectance minimum $\Delta(|S_{11}|^2)_{min}$ as a function of time since NaCl was added. The concentration values specify the NaCl solution after full dissolvement. (**E**) Reflectance $(|S_{11}|^2)_{min}$ and (**F**) frequency as a function of salinity for 500 µL NaCl solutions. The solid black lines show the fitting. The insets show the magnified plot in the range from 0 to 2 ppt.

**Discussion**

In this work, we have shown how the possibility of inducing a highly controllable localized BIC in a compact resonant structure by suitably tailoring the boundary conditions around it. Based on this concept, we have realized controllable on-demand symmetry breaking for advanced sensing. Specifically, we have exploited the natural attributes of rectangular waveguides to generate a single-resonator BIC whose coupling to the continuum of propagating modes in the waveguide can be controlled by a tiny drop of water. This approach overcomes the need for extreme materials or large-area geometries to induce BIC phenomena. Based on this platform, we have demonstrated the feasibility of the proposed system for sensing and to trace chemical reactions in real time by monitoring dissolution of NaCl in water and determining evaporation rates of distilled and salt water with a resolution of less than 1 μL. Although in our prototype design water was manually inserted with a pipette, the system can easily be extended to introduce a control system to manage the water insertion. Similar principles can be applied to more advanced microfluidic sensing platforms. In our design, the Q-factor of our BIC is limited by material losses, which can be reduced in different ways, including the use of better conducting materials and/or lower temperatures, further enhancing the overall sensitivity. Besides sensing, boundary-induced localized BICs can find applications in classical and quantum nanophotonic settings. For instance, the demonstrated enhanced sensitivity can be directly translated into enhanced light-matter interactions, offering a unique platform for drastically enhanced nonlinearities. In the context of qubits, these localized resonances may enable much enhanced quantum coherence times (*42*, *43*).

**REFERENCES AND NOTES**
1. J. von Neumann, E. P. Wigner, in *The Collected Works of Eugene Paul Wigner* (Springer Berlin Heidelberg, Berlin, Heidelberg, Heidelberg, 1993), pp. 291–293.
2. C. W. Hsu, B. Zhen, A. D. Stone, J. D. Joannopoulos, M. Soljacic, Bound states in the continuum. *Nat. Rev. Mater.* **1**, 16048 (2016).
3. A. Krasnok, D. Baranov, H. Li, M.-A. Miri, F. Monticone, A. Alú, Anomalies in light scattering. *Adv. Opt. Photonics*. **11**, 892 (2019).
4. F. Yesilkoy, E. R. Arvelo, Y. Jahani, M. Liu, A. Tittl, V. Cevher, Y. Kivshar, H. Altug, Ultrasensitive hyperspectral imaging and biodetection enabled by dielectric metasurfaces. *Nat. Photonics*. **13**, 390–396 (2019).
5. A. Kodigala, T. Lepetit, Q. Gu, B. Bahari, Y. Fainman, B. Kanté, Lasing action from photonic bound states in continuum. *Nature*. **541**, 196–199 (2017).
6. C. Huang, C. Zhang, S. Xiao, Y. Wang, Y. Fan, Y. Liu, N. Zhang, G. Qu, H. Ji, J. Han, L. Ge, Y. Kivshar, Q. Song, Ultrafast control of vortex microlasers. **1021**, 1018–1021 (2020).
7. S. T. Ha, Y. H. Fu, N. K. Emani, Z. Pan, R. M. Bakker, R. Paniagua-Domínguez, A. I. Kuznetsov, Directional lasing in resonant semiconductor nanoantenna arrays. *Nat. Nanotechnol.* **13**, 1042–1047 (2018).
8. M. Cotrufo, A. Alù, Excitation of single-photon embedded eigenstates in coupled cavity–atom systems. *Optica*. **6**, 799 (2019).
9. R. Duggan, Y. Ra'di, A. Alù, Temporally and Spatially Coherent Emission from Thermal Embedded Eigenstates. *ACS Photonics*. **6**, 2949–2956 (2019).
10. Z. Sakotic, A. Krasnok, N. Cselyuszka, N. Jankovic, A. Alù, Berreman Embedded Eigenstates for Narrow-Band Absorption and Thermal Emission. *Phys. Rev. Appl.* **13**, 064073 (2020).
11. Y. Ra'di, A. Krasnok, A. Alù, Virtual Critical Coupling. *ACS Photonics*. **7**, 1468–1475 (2020).
12. G. Gbur, Nonradiating sources and other invisible objects. *Prog. Opt.* **45**, 273 (2003).
13. A. Sommerfeld, Zur Elektronentheorie. I. Allgemeine Untersuchung des Feldes eines beliebig bewegten Elektrons. *Nachr. Akad. Wiss. Gottingen, Math.-Phys. Klasse*, 99–130 (1904).
14. A. Sommerfeld, Zur Elektronentheorie. II. Grundlagen fur eine allgemeine Dynamik des Elektrons. *Nachr. Akad. Wiss. Gottingen, Math.-Phys. Klasse*, 363–439 (1904).

15. P. Ehrenfest, Ungleichförmige Elektrizitätsbewegungen ohne Magnet- und Strahlungsfeld. *Phys. Zeitschrift*. **11**, 708–709 (1910).
16. G. A. Schott, The electromagnetic field of a moving uniformly and rigidly electrified sphere and its radiationless orbits. *London, Edinburgh, Dublin Philos. Mag. J. Sci.* **15**, 752–761 (1933).
17. F. Monticone, H. M. Doeleman, W. Den Hollander, A. F. Koenderink, A. Al, W. Den Hollander, A. F. Koenderink, A. Alù, Trapping Light in Plain Sight: Embedded Photonic Eigenstates in Zero-Index Metamaterials. *Laser Photon. Rev.* **12**, 1700220 (2018).
18. A. Krasnok, A. Alú, Embedded scattering eigenstates using resonant metasurfaces. *J. Opt.* **20**, 064002 (2018).
19. L. Vertchenko, R. Malureanu, C. DeVault, E. Mazur, A. V. Lavrinenko, in *Conference on Lasers and Electro-Optics* (OSA, Washington, D.C., 2020; https://www.osapublishing.org/abstract.cfm?URI=CLEO_SI-2020-SM1J.2), p. SM1J.2.
20. H. Friedrich, D. Wintgen, Physical realization of bound states in the continuum. *Phys. Rev. A*. **31**, 3964–3966 (1985).
21. C. W. Hsu, B. Zhen, J. Lee, S.-L. Chua, S. G. Johnson, J. D. Joannopoulos, M. Soljačić, Observation of trapped light within the radiation continuum. *Nature*. **499**, 188–191 (2013).
22. B. Zhen, C. W. Hsu, L. Lu, A. D. Stone, M. Soljačić, Topological Nature of Optical Bound States in the Continuum. *Phys. Rev. Lett.* **113**, 257401 (2014).
23. S. I. Azzam, V. M. Shalaev, A. Boltasseva, A. V. Kildishev, Formation of Bound States in the Continuum in Hybrid Plasmonic-Photonic Systems. *Phys. Rev. Lett.* **121**, 1–5 (2018).
24. J. Lee, B. Zhen, S. L. Chua, W. Qiu, J. D. Joannopoulos, M. Soljačić, O. Shapira, Observation and differentiation of unique high-Q optical resonances near zero wave vector in macroscopic photonic crystal slabs. *Phys. Rev. Lett.* **109**, 067401 (2012).
25. H. M. Doeleman, F. Monticone, W. Den Hollander, A. Alù, A. F. Koenderink, Experimental observation of a polarization vortex at an optical bound state in the continuum. *Nat. Photonics*. **12**, 397–401 (2018).
26. A. I. Ovcharenko, C. Blanchard, J.-P. Hugonin, C. Sauvan, Bound states in the continuum in symmetric and asymmetric photonic crystal slabs. *Phys. Rev. B*. **101**, 155303 (2020).
27. K. Koshelev, S. Lepeshov, M. Liu, A. Bogdanov, Y. Kivshar, Asymmetric Metasurfaces with High-Q Resonances Governed by Bound States in the Continuum. *Phys. Rev. Lett.* **121**, 193903 (2018).
28. C. A. Balanis, *Antenna theory: analysis and design* (New York ; Brisbane : J. Wiley, 1997).
29. S. Murai, D. R. Abujetas, G. W. Castellanos, J. A. Sánchez-Gil, F. Zhang, J. G. Rivas, Bound States in the Continuum in the Visible Emerging from out-of-Plane Magnetic Dipoles. *ACS Photonics* (2020), doi:10.1021/acsphotonics.0c00723.
30. E. N. Bulgakov, A. F. Sadreev, Bound states in the continuum in photonic waveguides inspired by defects. *Phys. Rev. B*. **78**, 075105 (2008).
31. Z. Liu, Y. Xu, Y. Lin, J. Xiang, T. Feng, Q. Cao, J. Li, S. Lan, J. Liu, High-Q Quasibound States in the Continuum for Nonlinear Metasurfaces. *Phys. Rev. Lett.* **123**, 253901 (2019).
32. F. Monticone, A. Alù, Embedded Photonic Eigenvalues in 3D Nanostructures. *Phys. Rev. Lett.* **112**, 213903 (2014).
33. S. Lannebère, M. G. Silveirinha, Optical meta-atom for localization of light with

quantized energy. *Nat. Commun.* **6**, 8766 (2015).
34. M. V. Rybin, K. L. Koshelev, Z. F. Sadrieva, K. B. Samusev, A. A. Bogdanov, M. F. Limonov, Y. S. Kivshar, High-Q Supercavity Modes in Subwavelength Dielectric Resonators. *Phys. Rev. Lett.* **119**, 243901 (2017).
35. A. M. Nicolson, G. F. Ross, Measurement of the Intrinsic Properties Of Materials by Time-Domain Techniques. *IEEE Trans. Instrum. Meas.* **19**, 377–382 (1970).
36. R. E. Jacobsen, A. V Lavrinenko, S. Arslanagic, Water-Based Metasurfaces for Effective Switching of Microwaves. *IEEE Antennas Wirel. Propag. Lett.* **17**, 571–574 (2018).
37. A. Andryieuski, S. M. Kuznetsova, S. V Zhukovsky, Y. S. Kivshar, A. V Lavrinenko, *Sci. Rep.*, in press, doi:10.1038/srep13535.
38. M. V Rybin, D. S. Filonov, K. B. Samusev, P. A. Belov, Y. S. Kivshar, M. F. Limonov, Phase diagram for the transition from photonic crystals to dielectric metamaterials. *Nat. Commun.* **6**, 10102 (2015).
39. W. J. Ellison, Permittivity of Pure Water, at Standard Atmospheric Pressure, over the Frequency Range 0–25THz and the Temperature Range 0–100°C. *J. Phys. Chem. Ref. Data*. **36**, 1–18 (2007).
40. A. Stogryn, Equations for Calculating the Dielectric Constant of Saline Water (Correspondence). *IEEE Trans. Microw. Theory Tech.* **19**, 733–736 (1971).
41. R. E. Jacobsen, A. V Lavrinenko, S. Arslanagic, A Water-Based Huygens Dielectric Resonator Antenna. *IEEE Open J. Antennas Propag.* **1**, 493–499 (2020).
42. M. Reagor, W. Pfaff, C. Axline, R. W. Heeres, N. Ofek, K. Sliwa, E. Holland, C. Wang, J. Blumoff, K. Chou, M. J. Hatridge, L. Frunzio, M. H. Devoret, L. Jiang, R. J. Schoelkopf, Quantum memory with millisecond coherence in circuit QED. *Phys. Rev. B*. **94**, 014506 (2016).
43. A. Blais, S. M. Girvin, W. D. Oliver, Quantum information processing and quantum optics with circuit quantum electrodynamics. *Nat. Phys.* **16**, 247–256 (2020).

**ACKNOWLEDGEMENTS**

The authors would like to thank the workshop at the Electromagnetic Systems group of the Technical University of Denmark for the fabrication of water-metal resonator elements for the experimental characterization.

**Funding:** This work was supported in part by the Simons Foundation and the Department of Defense.
**Authors contributions:** REJ, AK, AA conceived the idea. REJ designed the structure and carried out numerical and experimental work. REJ, AK, AA wrote the draft of the manuscript, which after several iterations, inputs and discussions of results with all authors, resulted in its final version.
**Competing interests:** The authors declare no competing interests.
**Data and materials availability:** All data are reported in the main paper and supplementary materials. Codes for the processing of data are given upon request.

**LIST OF SUPPLEMENTARY MATERIALS**
Materials and Methods
Fig. S1–S10

# Supplementary Materials to "Boundary-Induced Embedded Eigenstate in a Single Resonator for Advanced Sensing"

**AUTHORS:** Rasmus E. Jacobsen[1], Alex Krasnok[2], Samel Arslanagić[3], Andrei V. Lavrinenko[1*], and Andrea Alú[2,4*]

**AFFILIATIONS:**

[1]Department of Photonics Engineering, Technical University of Denmark, Bld. 345A, Ørsteds Plads, 2800 Kgs. Lyngby, Denmark (e-mail: rajac@fotonik.dtu.dk, alav@fotonik.dtu.dk)

[2]Photonics Initiative, Advanced Science Research Center, City University of New York, New York, NY 10031, USA (akrasnok@gc.cuny.edu, aalu@gc.cuny.edu)

[3]Department of Electrical Engineering, Technical University of Denmark, Bld. 348, Ørsteds Plads, 2800 Kgs. Lyngby, Denmark (e-mail: sar@elektro.dtu.dk)

[4]Physics Program, Graduate Center, City University of New York, New York, NY 10016, USA

*Corresponding author: e-mail: alav@fotonik.dtu.dk, aalu@gc.cuny.edu

**This PDF file includes:**

    Materials and Methods
    Supplementary Text
    Figs. S1 to S10

**Materials and Methods**
Simulations
COMSOL Multiphysics 5.5 was used for all numerical simulations. The Electromagnetic Waves, Frequency Domain physics, was used with the Frequency Domain and Eigenfrequency studies. The permittivity model for water was implemented (see equations in the next section). Impedance boundary conditions with finite intrinsic conductivity were applied on the metallic surfaces. The magnitude of the incidence wave is 1 V/m in all simulations. Two models were used (sketches are shown in Fig. S10):

1. **2-D array:** Periodic boundary conditions were applied to emulate the infinite extent of the array. A periodic port backed with PML and plane wave excitation was used to retrieve the $S_{11}$ scattering parameter. Intrinsic conductivity of $5.8\times10^7$ S/m (corresponding to copper) were applied on the metallic surfaces.
2. **Single resonator in WR-430 rectangular waveguide:** Perfect electric conductors (PECs) were applied on the metallic walls of the waveguide. A rectangular port backed with Perfect Matched Layers (PML) and $TE_{10}$ excitation was used to retrieve the $S_{11}$ scattering parameter. Intrinsic conductivity of 3.8×10$^7$ S/m (corresponding to aluminum) were applied on the metallic surfaces of the resonator and conducting ground plane. Permittivity of 1.075 is applied for the Rohacell 51 HF material.

Permittivity model for water
The permittivity for distilled water was calculated using the following equations. The relative permittivity as a function of temperature $T_w$ [°C] and angular frequency $\omega$ [rad/s] is (*39*)

$$\varepsilon_{r,w}(\omega, T_w) = \varepsilon'_{r,w} - j\varepsilon''_{r,w} = \varepsilon_\infty(T_w) + \frac{\varepsilon_s(T) - \varepsilon_\infty(T_w)}{1 - j\omega\tau(T_w)} \tag{1}$$

with $\varepsilon_s(T_w) = a_1 - b_1 T_w + c_1 T_w^2 - d_1 T_w^3$ and $\varepsilon_\infty(T_w) = \varepsilon_s(T_w) - a_2 \exp(-b_2 T_w)$ being the optical and static permittivities, respectively. $\tau(T_w) = \tau_0 \exp[T_1/(T_w - T_0)]$ is the rotational relaxation time. The values for the constants are $a_1 = 87.9$, $b_1 = 0.404$ K$^{-1}$, $c_1 = 9.59 \times 10^{-4}$ K$^{-2}$, $d_1 = 1.33 \times 10^{-6}$ K$^{-3}$, $a_2 = 80.7$, $b_2 = 4.42 \times 10^{-3}$ $K^{-1}$, $\tau_0 = 0.137$ ps, $T_1 = 651$ °C and $T_0 = 133$ °C.

Experiment with a single resonator in the rectangular waveguide
The resonator was fabricated at the local workshop and is made of an aluminum cylinder mounted on an aluminum plate with screws on the back. Holes were drilled on the aluminum plate for the attachment to the rectangular waveguide. The water container was made by drilling a hole in a small cylindrical Rohacell 51 HF block, which was then glued on the aluminum cylinder with glue.

The standard WR-430 rectangular waveguide is made of brass and has a cross-section of 109.22 × 54.61 mm$^2$. A WR-430 flange waveguide to coax adapter was attached to it, which was connected to Anritsu MS2024B Vector Network Analyzer (VNA) with a 1 m RG58 50 Ohm coaxial cable. The VNA was calibrated for S$_{11,\text{VNA}}$ magnitude response using a short (brass plate). To retrieve the phase change as well as to remove the oscillations caused by the Fabry–Pérot interferometer effects, we defined the scattering parameter as S$_{11}$ = S$_{11,\text{VNA}}$($V_{\text{water}}$)/ S$_{11,\text{VNA}}$(0 µL), where $V_{\text{water}}$ is the volume of water. See Fig. 4C for sketch of the experimental setup.

Local distilled water was used, and an Eppendorf Research 10–200 µL pipette was used to insert the liquids. The solid NaCl crystals were weighted with a scale at the local laboratory. The NaCl solution was made using a weight scale and by mixing the distilled water with solid NaCl crystals. The water and the NaCl solutions were stored together in closed containers to ensure similar temperatures. For heating of water, we used a simple boiler.

The measurements were done with the following steps
1. Insert liquid.
2. Close waveguide.
3. Wait for the stable response due to mechanical vibrations and record the S$_{11}$-response.
4. Prepare for the next measurement by removing the waveguide and dry for liquid.

Normalization
The extracted reflection coefficient oscillates around 1, which is unphysical, and must be an artifact coming from the VNA calibration. Such oscillations have been observed previously with other structures inserted in waveguides (*36*). The waveguide is a closed cavity and inside the waves reflects multiple times, causing interference just like in Fabry–Pérot interferometers. The oscillations in Fig. S5A occur due to a shift in the interfering reflections. Since we are not interested in these oscillations, we suppress them by defining a normalized reflection coefficient as $S_{11} = S_{11,\text{VNA}}(V_{\text{water}})/ S_{11,\text{VNA}}(0\ \mu\text{L})$. The magnitude and phase of $S_{11}$ are shown in Fig. S5, B and C, respectively, as functions of frequency for different volumes of water.

Quality factor
The quality factor has been calculated using $Q = f_r/\Delta f$, where $f_r$ is the resonance frequency and $\Delta f$ is the bandwidth at half maximum. The quality factor as a function of the volume of water is shown in Fig. S5E. We observe a decreasing linear proportionality of the quality factor with an increasing volume of water. This is expected since water is lossy. At critical coupling, the quality factor is around 250.

Reproducability of the experiment
We tested the reproducability of the experiment, see Fig. S5F, and we measured a variation in the resonance frequency (the frequency of the minimum of the reflection coefficient) of approximately ±0.008 % (±0.15 MHz), whereas the reflection magnitude had a variation around ±4 %. These deviations come from small variations in water (shape and volume) and are due to the manual insertion of the water.

It should be noted that the resolution is profoundly dependent on the accuracy and precision of the water insertion. Although in our prototype, the water was manually inserted with a pipette, it can be improved with a control system to manage the water insertion.

Water outside of the Rohacell container
We also studied placing the water outside the Rohacell container, and we observe similar behavior for the volume change. These results are shown in Fig. S8 and S9. As the water can move freely on the metallic cylinder, we get far more variations in the measurements, and therefore the measurements are more difficult to replicate, but at the same time, makes our device far more sensitive to movements. The position of water is very important, and only displacement parallel to the incident electric field will excite the resonance.

Supplementary figures
See next pages.

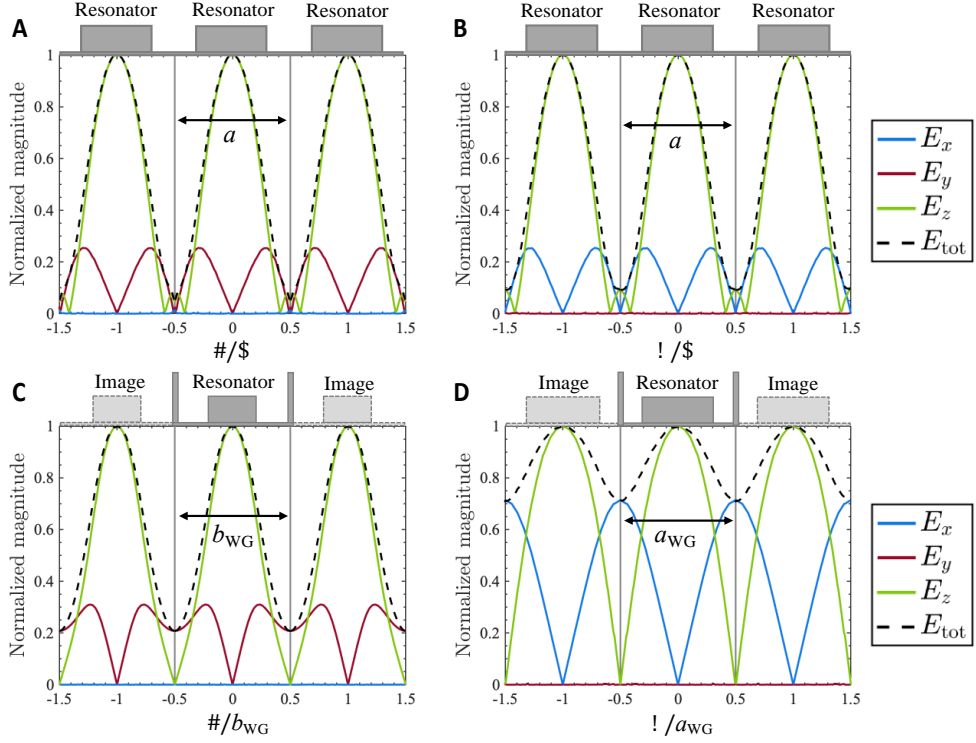

**Figure S1. Difference between the rectangular waveguide with perfect electric conductor (PEC) walls and the periodic array (TE$_{10}$ mode off).** Magnitude of the local E-field components normalized with the total E-field ($E_{\text{tot}}$) at $z = h_{\text{cyl}} - 20$ mm for the eigenmode analyses. (**A**) and (**B**) Periodic array: $f_r \approx 2470$ MHz, $r = 30$ mm, $h = 25$ mm and $a = 110$ mm. (**C**) and (**D**) Rectangular waveguide with PEC walls: $f_r \approx 1890$ MHz, $r = 18$ mm, $h = 27$ mm and $a_{\text{WG}} = b_{\text{WG}}/2 = 54.61$ mm. In (**A**) and (**C**), $x = 0$, whereas $y = 0$ in (**B**) and (**D**). The $z$-component of the E-field is the most pronounced in both configurations. The $z$-component of the images in (**B**) and (**D**) lags 180° due to the PEC walls of the waveguide, see e.g. Fig. S3.

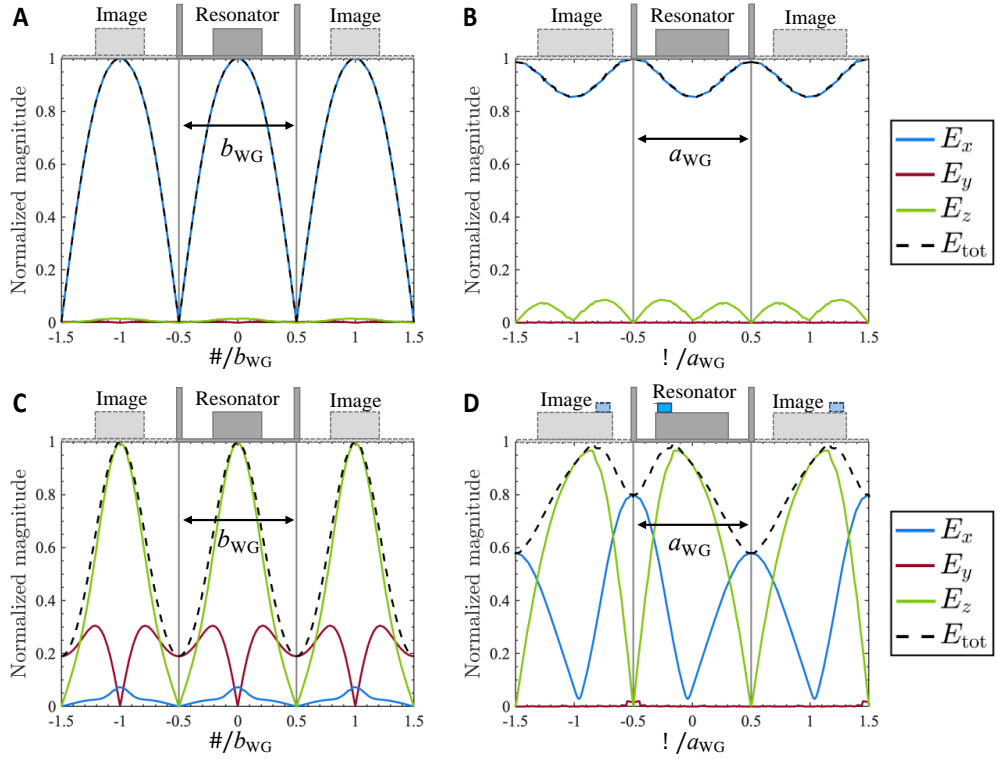

**Figure S2. Rectangular waveguide of perfect electric conductor (PEC) walls with and without water (TE$_{10}$ mode on).** Magnitude of the local E-field components normalized with the total E-field ($E_{tot}$) at $z = h_{cyl} - 20$ mm for the TE$_{10}$ mode excitation at resonance frequency. The geometrical parameters are $r = 18$ mm, $h = 27$ mm and $a_{WG} = b_{WG}/2 = 54.61$ mm. (**A**) and (**B**) Without water at $f_r \approx 1890$ MHz. (**C**) and (**D**) With 500 µL water at $f_r \approx 1830$ MHz. In (**A**) and (**C**), $x = 0$, whereas $y = 0$ in (**B**) and (**D**). The $z$-component of the E-field is nearly absent without water as we do not excite the resonance. With water, there is a large z-component since the resonance is know excited. The $z$-component of the images in (**B**) and (**D**) lags 180° due to the PEC walls of the waveguide, see Fig. S3.

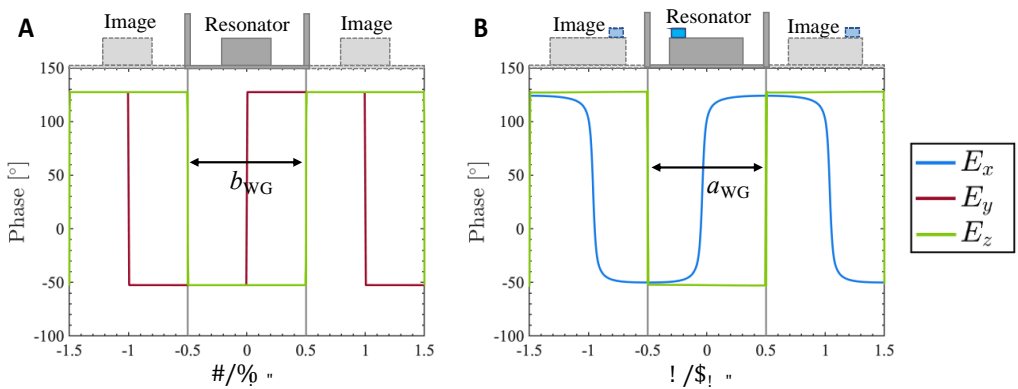

**Figure S3. Same configuration as in Fig. S2C and D, but showing phase instead.** Phase of the local electric field components at $z = h_{cyl} - 20$ mm. In (**A**) $x = 0$, whereas $y = 0$ in (**B**). The $z$-component of the images lags 180° due to the PEC walls of the waveguide.

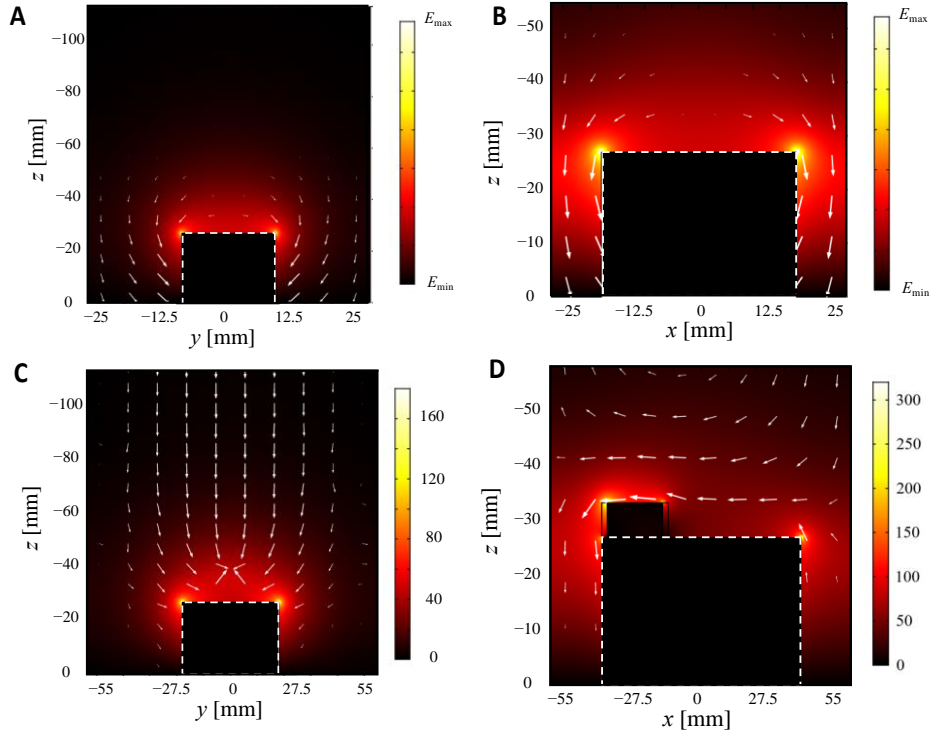

**Figure S4. Total E-field magnitude in the rectangular waveguide with the metallic resonator.** (**A**) and (**B**) Eigenmode analysis of the resonator in the waveguide. The real eigenfrequency is 1878.9 MHz and Q-factor is 8,300. (**C**) and (**D**) $TE_{10}$ mode wave excitation with 500 µL of destilled water in a cylindrical Rohacell 51 HF container of height 6.4 mm as well as inner and outer radii of 5 mm and 6 mm, respectively, atop of the resonator. The edge of Rohacell 51 HF container coincides with the edge of the resonator. The frequency is 1829.1 MHz, the Q-factor is 600 and the reflectance is 0.06. (**A**)-(**D**) Colorbar shows the total E-field magnitude in V/m. Arrows show the power flow density in logarithmic scale. The radius and height of the metallic resonator are $r = 18$ mm and $h = 27$ mm, respectively. The material of the metallic resonator and conducting ground plane is aluminum with intrinsic conductivity $\sigma = 3.8 \times 10^7$ S/m. The walls of the waveguide are PEC. The temperature of water is 20 °C in all simulations.

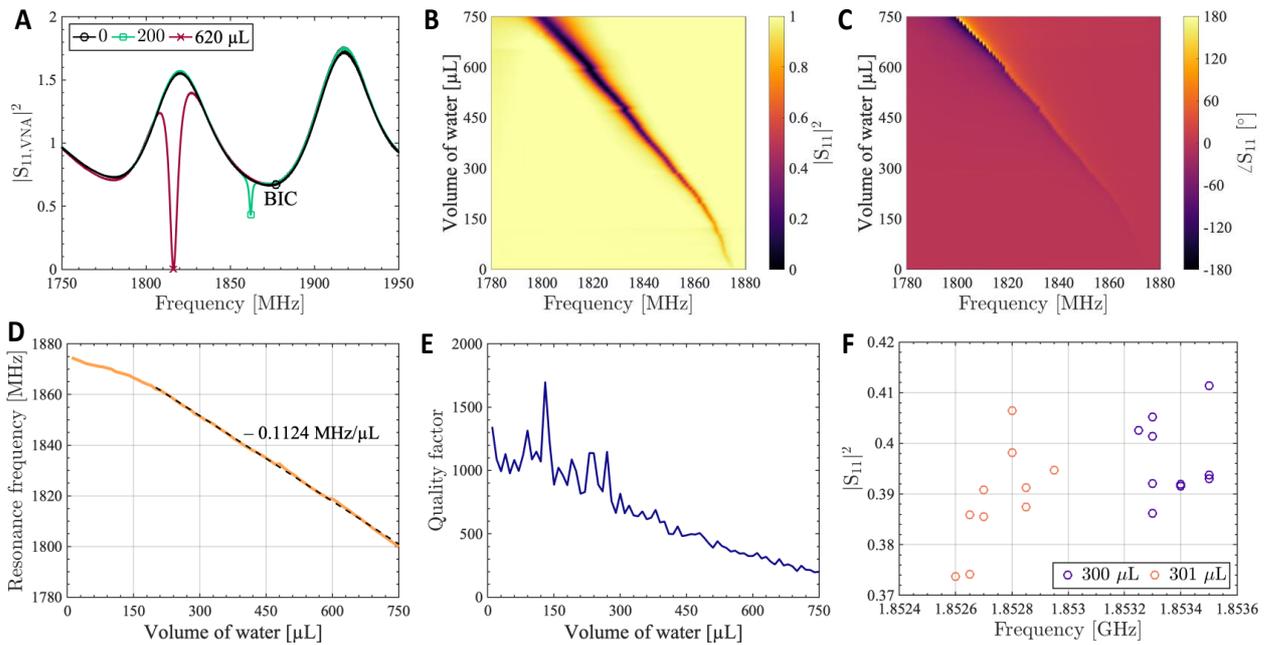

**Figure S5. Supplementary figures to Fig. 4.** (**A**) Measured reflection coefficient $S_{11,\text{VNA}}$ as a function of frequency for different volumes of water inserted in the Rohacell container. (**B**) and (**C**) Magnitude and phase, respectively, of the normalized reflection coefficient $S_{11}$ as a function of frequency and volume of water. (**D**) Resonance frequency as a function of the volume of water. The dashed line shows the tendency from 200 µL. (**E**) Caculated quality factor as a function of the volume of water. (**F**) Repeatability of measurements. Reflectance $|S_{11}|^2$ as a function of frequency for 300 µL and 301 µL of water. The measurements are repeated 10 times for both volumes

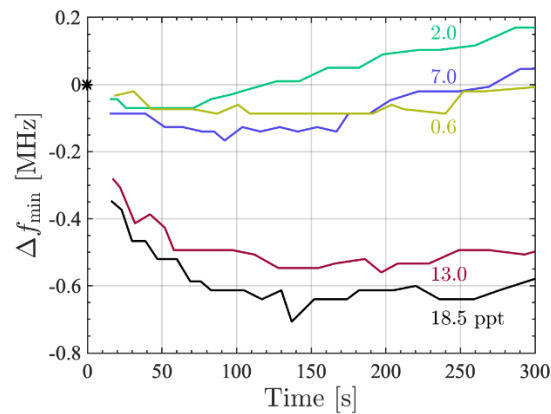

**Figure S6. Supplementary figure to Fig. 5.** Frequency change ($\Delta f_{\min}$) in the reflectance minimum $(|S_{11}|^2)_{\min}$ as a function of time since NaCl was added. The asterix symbol shows the frequency of a distilled water (0 ppt).

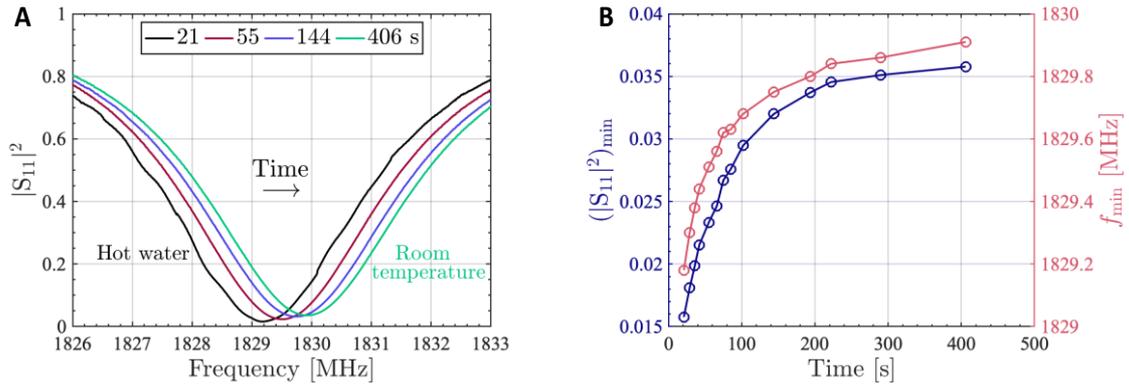

**Figure S7. Change in water temperature.** Initial volume of water is 500 μL for all experiments. (**A**) reflectance as a function of frequency for different time points after adding the heated water. (**B**) The minimum of the reflectance $(|S_{11}|^2)_{min}$ and the frequency $f_{min}$ as functions of the time after adding the heated water.

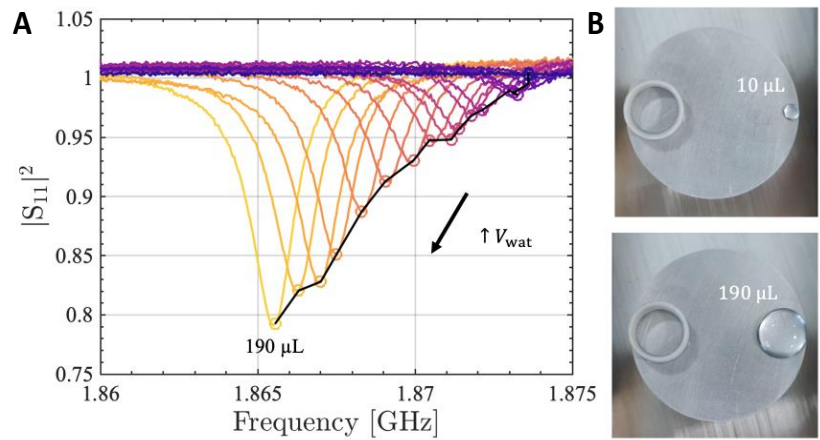

**Figure S8. Water drop outside of the Rohacell container.** (**A**) Reflectance $|S_{11}|^2$ as a function of frequency for different volumes of water. (**B**) Photographs showing 10 μL and 190 μL of water placed on the resonator.

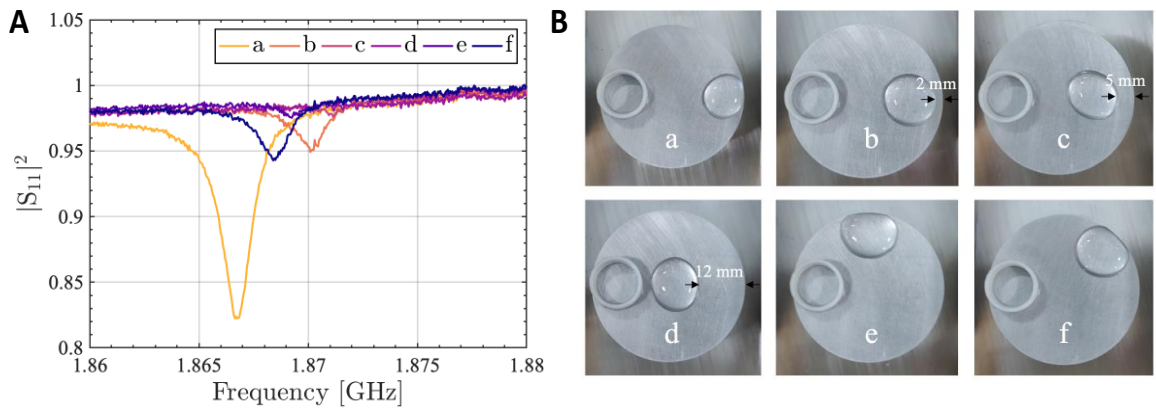

**Figure S9. Water drop of 200 µL outside of the Rohacell container placed at different positions on the the resonator.** (**A**) Reflectance $|S_{11}|^2$ as a function of frequency for different water drop positions. (**B**) Photographs showing the positions of the drop on the resonator.

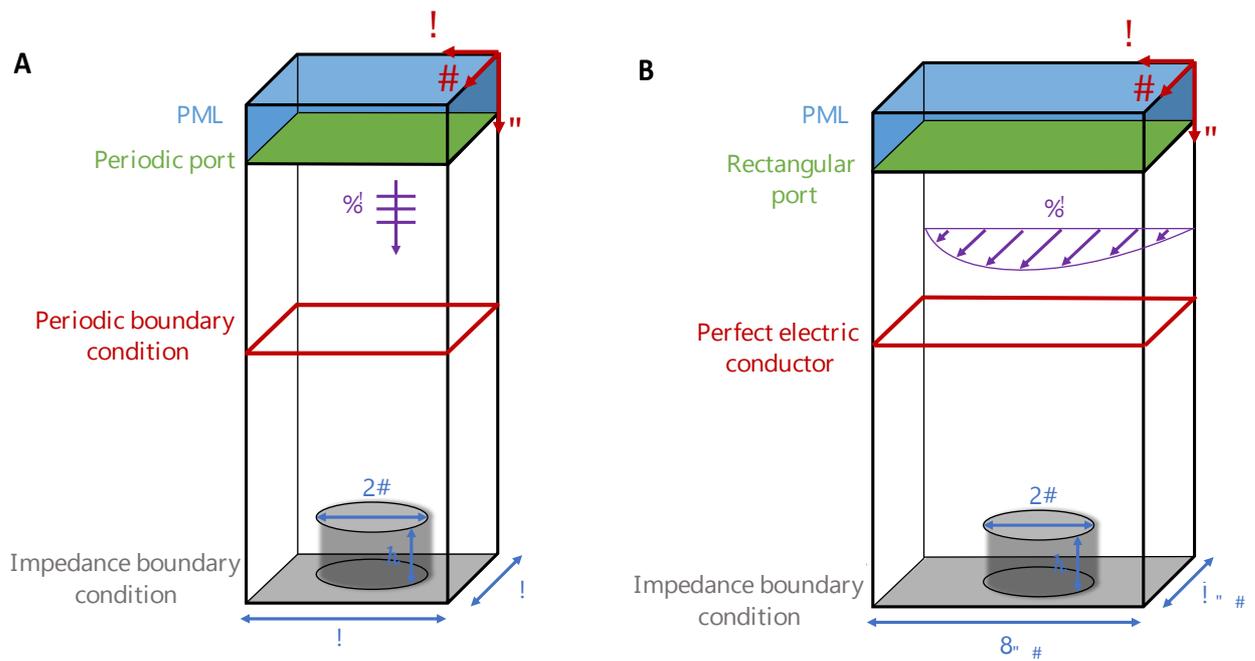

**Figure S10. Sketches of the COMSOL Multiphysics simulation models.** (**A**) Model of the BIC structure with infinite elements. (**B**) Model of the rectangular waveguide used for the experiments.